# Quantum propensity in economics

David Orrell[1]

Monireh Houshmand[2]

**Abstract**: This paper describes an approach to economics that is inspired by quantum computing, and is motivated by the need to develop a consistent quantum mathematical framework for economics. The traditional neoclassical approach assumes that rational utility-optimisers drive market prices to a stable equilibrium, subject to external perturbations. While this approach has been highly influential, it has come under increasing criticism following the financial crisis of 2007/8. The quantum approach, in contrast, is inherently probabilistic and dynamic. Decision-makers are described, not by a utility function, but by a propensity function which specifies the probability of transacting. We show how a number of cognitive phenomena such as preference reversal and the disjunction effect can be modelled by using a simple quantum circuit to generate an appropriate propensity function. Conversely, a general propensity function can be quantized to incorporate effects such as interference and entanglement that characterise human decision-making. Applications to some common problems in economics and finance are discussed.

## 1. Introduction

Theories of economics always rely on theories of value. In classical economics, it was assumed that value is the product of labour (Smith, 1776). A problem of course is that labour is difficult to measure, and the link with value is not always clear. This became obvious with the development of sophisticated financial markets in the late nineteenth century, where prices for something like a share in a railway company didn't directly reflect human labour. Neoclassical economists reacted by substituting labour with utility, which had been defined by Bentham in his 1789 *Principles of Morals and Legislation* as that which appears to "augment or diminish the happiness of the party whose interest is in question."

---

[1] Systems Forecasting, Toronto, Canada
[2] Department of Electrical Engineering, Imam Reza International University, Mashhad, Iran. m.hooshmand@imamreza.ac.ir.

As with labour, utility seemed hard to measure directly, but as Jevons pointed out in his *Theory of Political Economy* (1871), another approach was to simply assume that utility was reflected by price. While this turned utility into a somewhat circular concept (Robinson, 1962), it did seem to hold out the possibility that economics could be put on a rational scientific basis; a task eventually undertaken by Von Neumann and Morgenstern, whose *Theory of Games and Economic Behaviour* (1944) developed a consistent set of axioms to describe rational economic behaviour based on a theory of expected utility.

The assumption that people act rationally to optimise their own utility became the basis for economics as it developed in the post-war era. However while this model of rational economic behaviour remains the default approach in economics, cognitive psychologists have shown that its assumptions are often violated. For example, one of the key axioms of expected utility theory is that people have fixed preferences. Yet the widely-demonstrated phenomenon of preference reversal shows that in fact people do not have stable preferences and have a tendency to change their mind depending on things like context (Tversky and Thaler, 1990).

The belief that rational utility-optimisers drive prices to a stable equilibrium was also sorely tested by the financial crisis of 2007/8. In response to that crisis, economists began to adopt methods from behavioural economics, in which so-called cognitive anomalies were accommodated by modifying the utility function to account for effects such as loss aversion or herd behaviour (Kahneman, 2011). As discussed below, though, a range of cognitive phenomena continue to elude behavioural approaches, because they do not conform to classical logic (Wendt, 2015). This has motivated interest in adopting a mathematical framework based on quantum probability.

Quantum probability is a set of mathematical rules to calculate probabilities of events in quantum mechanics. Its properties such as nonadditivity and noncommutativity make it well-suited to model uncertainty in decision-making behavior in social sciences, particularly in economics and more generally human cognition where it extends classical utility theory (Qadir, 1978; Segal, 1998; Baaquie, 2004; Derman, 2004; Busemeyer and Bruza, 2012; Haven and Khrennikov, 2013). The approach known as quantum decision theory, first proposed in (Yukalov and Sornette, 2008), provides a framework for modeling and predicting human decisions, that can be used to explain many behavioural phenomena which are unaccounted by utility-theoretic approaches. Quantum decision theory allows us to characterize the rational–irrational dual behavior of real decision makers, who evaluate not only the utility of the given prospects, but also their respective attractiveness. In (Yukalov and Sornette, 2014) and (Yukalov and Sornette, 2015), the lotteries consisting only of gains are considered, then the idea is extended to lotteries with both gain and loss (Yukalov and Sornette, 2017). In (Zhang, C. and Kjellström, 2021) multi-modal attraction factor function that take multiple emotional and cognitive items into account are considered.

The approach in this paper is to generalize the quantum approach, by modelling economic decision-making using a probabilistic propensity function that can be expressed in quantum terms. In other words, we model the economy in a manner consistent with, and inspired, by quantum computing (Nielsen and Chuang, 2002; Nakahara M and Ohmi T, 2008). The plan of the remainder of the paper is as follows. Section 2 motivates the use of quantum probability to model economic decisions. Section 3 shows how a simple quantum circuit, of a sort commonly used in quantum algorithms, can simulate a variety of cognitive phenomena which elude a classical approach. Section 4 shows how a general propensity function can be quantized to yield the quantum dynamics of financial transactions. Finally Section 5 summarises the main results.

## 2. Quantum probability

The key difference between quantum computers and classical computers is that, instead of storing information in binary bits which can take on the value 0 and 1, quantum computers use qubits, which randomly collapse to a particular state when measured. Quantum computers are therefore inherently probabilistic rather than deterministic, so a quantum circuit may have to be run and measured many times in order to build up a statistical estimate to a solution. Probabilities of quantum measurements are modelled by a 2-norm (which involves a sum of squares) rather than a 1-norm.

In order to motivate, from first principles, the use of quantum probability in a social context, suppose that we wish to model the state of a person who is going to make a binary choice between two options. If the person is equally likely to choose either, then the situation is the same as for a random coin toss. In general, the state could be modelled by the diagonal ray in Figure 1.A, where two dimensions are required because we want to capture the possibility of either outcome. If the ray has length 1, and we associate the probabilities of obtaining heads or tails by taking the square, i.e. the 2-norm, of the projections onto the respective axis, then the probabilities add to 1 as expected.

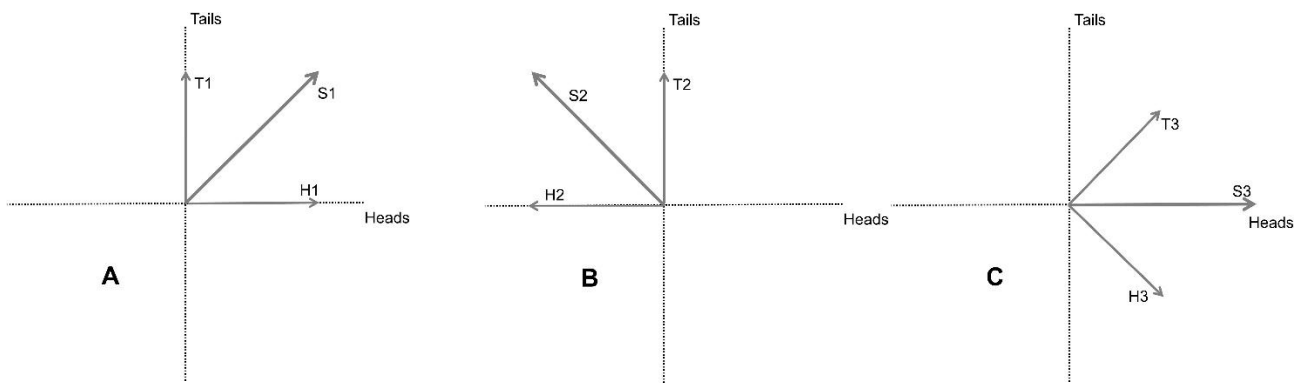

Figure 1. A) A coin toss for a balanced coin can be expressed as a superposition of two states, heads and tails. B) because the 2-norm of a probability is its square, we can also consider negative projections. C) Applying the Hadamard transformation rotates S1 by 45 degrees clockwise which aligns with the H axis (S3). The ray H3 (the rotated version of H1) can be viewed as a superposition of horizontal and vertical parts (not shown), and the same for T3 (the rotated version of T1); however the vertical parts now cancel out, leaving only a horizontal ray S3.

The state can therefore be interpreted as a propensity to give different outcomes, in this case heads or tails. Because the 2-norm of a probability depends on the square, one can also imagine cases where the projections are negative (Haug, 2004; Aaronson, 2013). For example, in Figure 1.B the projection on the heads axis is negative, but the 2-norm of the probability is unchanged. Since negative numbers are allowed, the need for mathematical closure suggests that complex numbers should be as well, for example to accommodate situations where we need to calculate square roots (Aaronson, 2013).

Suppose now that we apply a unitary (and so norm-preserving) transformation, known as the Hadamard transformation, which rotates S1 by 45 degrees clockwise. A coin in the superposed state of Figure 1.A will then be rotated so that it aligns with the H axis, as in Figure 1.C, which means that when measured it will be Heads for sure. The indeterminate system has become deterministic. Another way to think about this is to recall that the initial propensity state is a superposition of two orthogonal states, denoted H1 and T1 in the figure. After being rotated, T1 leads to T3 and H1 leads to H3. The T3 ray has a positive heads and a positive tails component. However the H3 ray has a positive heads and a negative tails. The positive and negative tails components therefore cancel out, leaving only the positive heads components. This is an example of interference, which again occurs because the 2-norm allows negative terms.

Finally, we can also ask what happens when we have a more complicated system, for example two coins instead of one. The possible final outcomes are then HH, HT, TH, and TT, where HT is heads for the first coin and tails for the second, and so on. But now, representing the superposed state of the system is considerably more complicated, because instead of just needing two axes, H and T, we now need four axes: HH, HT, TH and TT. For some situations, it is possible to treat each coin separately, which would allow us to represent our information about each one in its own two-dimensional figure; but if say the system is in a balanced superposition of HH and TT – i.e. there is an equal probability of getting either both heads, or both tails, and those are the only possibilities – then we say that the coins are entangled. Entanglement can therefore be viewed as a particular kind of superposed state, which leads to correlation in measurement probabilities of subsystems.

The adoption of the 2-norm as a probabilistic measure therefore models the related phenomena of superposition, interference, and entanglement, which are characteristic of quantum systems, and are exploited in quantum computers to give significant computational advantages over classical computers (Aaronson, 2013). In a quantum computer, the coin toss would be modelled using the wave function $|\psi\rangle$ of a single qubit, which is in a superposition of two basis states $|0\rangle = \begin{pmatrix}1\\0\end{pmatrix}$ and $|1\rangle = \begin{pmatrix}0\\1\end{pmatrix}$, with the first representing heads and the second representing tails. We can then write

$$|\psi\rangle = a_0|0\rangle + a_1|1\rangle = \begin{pmatrix}a_0\\a_1\end{pmatrix}$$

where $a_0$ and $a_1$ are complex numbers with $|a_0|^2 + |a_1|^2 = 1$. When measured in the computational basis, the possible outcomes are $|0\rangle$ with a probability $|a_0|^2$ and $|1\rangle$ with a probability $|a_1|^2$. Quantum gates which produce transformations are represented by unitary matrices, and two or more qubits are represented by tensor products, as seen below.

To summarise, classical probability is the simplest kind of probability, which is based on the 1-norm and involves positive numbers. The next-simplest kind of probability uses the 2-norm, and includes complex numbers. The reason this kind of probability is called quantum probability, is for the historical reason that it turns out to be the right framework for quantum physics, and is the basis of quantum computing; however there is no reason we can't apply it to other areas, such as economics.

## 3. Quantum circuits

Quantum probability has been adopted in areas such as quantum cognition and quantum game theory because it naturally accounts for effects such as interference and entanglement (Busemeyer and Bruza, 2012; Wendt, 2015; Haven et al, 2018). Well-known examples from the quantum cognition literature include the order effect, where responses to questions in a survey depend on the order in which they are asked; the disjunction effect, where extra information seems to interfere with decision-making in a manner that eludes classical logic; preference reversal, where a decision changes depending on context; or games such as the prisoner's dilemma, where experiments show that people behave not as individual utility-optimisers, but as people who are entangled through things like a social contract.

For the order effect, the usual way to think about this is in terms of a sequence of projections, as illustrated in Figure 2. If question A is followed by question B, and we assume that the questions have yes/no responses, then the first response is modelled by collapsing the state, shown by the diagonal grey line, onto one of the two axes labelled "A yes" or "A no". That state is then used as the starting point for a projection onto the B axes. The probabilities of the possible outcomes when question A is asked first are as shown in Table 1. Similar calculations can be made with the order reversed to reveal the order effect, which has been demonstrated in a broad range of empirical studies (Wang et al, 2014). The probability of outcomes when question A is followed by question B and vice versa are shown in Table 2.

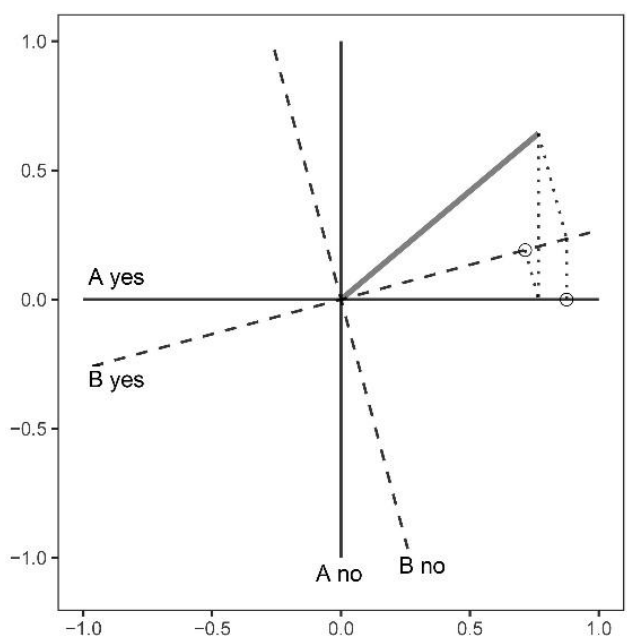

Figure 2. The order effect for two questions labelled A and B. The state vector (grey line) is at an angle $\theta$ to the axes for A. The axes for B are rotated by an angle $\varphi$ to those for A.

Table 1. Probabilistic outcomes from the quantum circuit for a sequence of two queries A and then B.

| Measured state | Response A | Probability | Response B | Conditional Probability | Joint Probability |
|---|---|---|---|---|---|
| $|00\rangle$ | Yes | $\cos^2\theta$ | Yes | $\cos^2\varphi$ | $\cos^2\theta \cos^2\varphi$ |
| $|01\rangle$ | Yes | $\cos^2\theta$ | No | $\sin^2\varphi$ | $\cos^2\theta \sin^2\varphi$ |
| $|10\rangle$ | No | $\sin^2\theta$ | Yes | $\sin^2\varphi$ | $\sin^2\theta \sin^2\varphi$ |
| $|11\rangle$ | No | $\sin^2\theta$ | No | $\cos^2\varphi$ | $\sin^2\theta \cos^2\varphi$ |

Table 2: Probabilities of the possible outcomes based on order effect

| Question order | A yes | A no | B yes | B no |
|---|---|---|---|---|
| A then B | $\cos^2\theta$ | $\sin^2\theta$ | $\cos^2\theta \cos^2\varphi + \sin^2\theta \sin^2\varphi$ | $\cos^2\theta \sin^2\varphi + \sin^2\theta \cos^2\varphi$ |
| B then A | $\cos^2(\theta-\varphi)\cos^2\varphi + \sin^2(\theta-\varphi)\sin^2\varphi$ | $\cos^2(\theta-\varphi)\sin^2\varphi + \sin^2(\theta-\varphi)\cos^2\varphi$ | $\cos^2(\theta-\varphi)$ | $\sin^2(\theta-\varphi)$ |

An equivalent representation of this sequence, based on the methods of quantum computing, is the quantum circuit shown in Figure 3. The input on the left is two qubits, each of which is initialised to $|0\rangle$. The joint initial state is written

$$\psi_0 = |00\rangle = \begin{pmatrix}1\\0\end{pmatrix} \otimes \begin{pmatrix}1\\0\end{pmatrix} = \begin{pmatrix}1\\0\\0\\0\end{pmatrix}.$$

The top qubit is acted on by the gate $A$, which is a rotation matrix of the form

$$A = R_\theta = \begin{pmatrix}\cos\theta & -\sin\theta\\ \sin\theta & \cos\theta\end{pmatrix}.$$

The lower qubit is acted on by the similar rotation matrix $B = R_\varphi$ where $\varphi$ represents the difference in the frameworks used to answer the two questions. The two qubits are then entangled through a C-NOT gate, which is represented by the matrix

$$X_c = \begin{pmatrix}1 & 0 & 0 & 0\\ 0 & 1 & 0 & 0\\ 0 & 0 & 0 & 1\\ 0 & 0 & 1 & 0\end{pmatrix}.$$

The operation $\psi_f = X_c(A \otimes B)\psi_0$ depicted in Figure 3 then yields the same output probabilities for the possible states as in Table 1. Similar calculations can be made with the order of questions reversed, by inverting the C-NOT gate so that the second qubit acts as the control and updating $A$ as $R_\varphi$ and $B$ as $R_{(\theta-\varphi)}$.

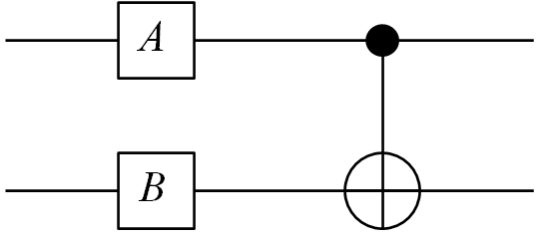

Figure 3. Quantum circuit for a decision $B$ influenced by a context $A$.

As seen in the Appendix, this relationship holds when gates $A$ and $B$ are arbitrary unitary matrices, so the circuit is quite flexible and can be used to simulate any problem that can be described either by a sequence of projections, as in quantum cognition, or by an entangled system as in quantum decision theory (Yukalov and Sornette, 2015). For the disjunction effect (Blutner and Graben, 2016), gate $A$ represents a context or a piece of information, while gate $B$ represents a decision. For preference reversal (Yukalov and Sornette, 2015), gate $A$ represents a subjective context, while gate $B$ represents an objective utility such as a monetary award. The circuit can also be used to simulate a version of the prisoner's dilemma, where gate $B$ represents one player's strategy, and gate $A$ represents their ideas about the other player's strategy (Orrell, 2020b).

A particularly strong, and economically relevant, example of a quantum social phenomenon is the existence of threshold effects. For the case of preference reversal, if the context-dependent subjective factors represented by gate $A$ are assumed to be random, then they can be expected to have a roughly equal effect as the objective factors (such as the prize in a lottery) represented by gate $B$. Because changes in context often have a switch-like nature (for example, a person may or may not have a particular piece of information or experience a particular event) the result is a kind of threshold effect, where objective costs are balanced by subjective factors which can suddenly change. According to the preference reversal criterion (Yukalov and Sornette, 2018), if the less attractive option has an associated cost $x_1$ and the more attractive option has a cost $x_2$, then a switch from the more attractive to the less attractive option will only occur if $x_2/x_1 > 3$. This criterion has been empirically tested in a range of experiments, and appears to be quite robust (Yukalov and Sornette, 2015; 2018; Orrell, 2021a). We return to the question of threshold effects below.

## 4. Propensity and entropic force

The above examples show that the quantum approach, native to quantum computers, is well-suited to studying a variety of problems that involve interference and entanglement, and are therefore not easily addressed using classical logic or behavioural approaches, which is why quantum methods are seeing increasing use in the social sciences (Wendt, 2015; Der Derian and Wendt, 2020). For the case of economics the quantum approach, with its change from utility to propensity, also leads to a shift in our understanding of how decisions are made, and how financial transactions are modelled.

Smith (1776) argued that the "propensity to truck, barter, and exchange" was inherent in human nature – and while we can't directly observe utility, we can certainly observe people buying and selling. We can therefore define a propensity function as a kind of schedule which describes the probability that a person will take a certain decision. Similar propensity curves are used in marketing, where the technique known as "propensity modelling" is used to simulate how a customer's willingness to buy is affected by attributes including price (Wilcox, n.d.). Since as we have seen quantum probability can be used to derive a propensity function for the discrete case, a natural question to ask is whether it is possible to go the other way, and use a given propensity function to derive a quantum model.

As an example, suppose that we made a probabilistic estimate of the price of a house. The result could resemble a normal distribution, as in Figure 4, where the center of the distribution would be our best estimate, while the standard deviation would be a measure of our uncertainty. In quantum terms, this can be interpreted as a superposition state, where the chance of selecting a particular price depends on the squared amplitude of an underlying wave function. The situation is therefore similar to the coin toss, except that instead of only heads or tails there is now a continuous range of possible outcomes.

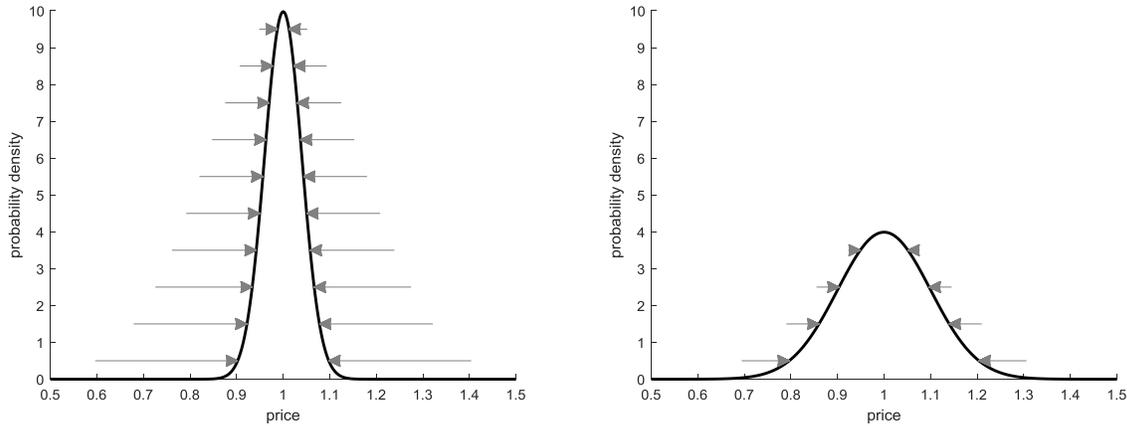

Figure 4. The curves show propensity as a function of price, measured in millions of dollars. Both are centered at $P = 1$, but the panel on the right has a higher level of price flexibility. The arrows indicate the strength and direction of the associated entropic forces (discussed later).

While it is traditional in economics to talk about the forces of supply and demand, these forces are assumed to cancel at equilibrium, and there is no consistent concept of economic mass. The dynamics of economic transactions are therefore not usually considered in detail, other than to assume the system is at balance. In contrast, an advantage of the propensity framework is that it leads to the concept of entropic force, which reflects the tendency of a system to achieve maximum entropy (Sokolov, 2010; Caticha, 2019). In statistical physics, an entropic force for a probability distribution $P(x)$ is given by

$$f(x) = \gamma \frac{P'(x)}{P(x)} = k_B T \frac{P'(x)}{P(x)}$$

where $T$ is temperature and $k_B$ is the Boltzmann constant. In economics, the propensity can similarly be viewed as the product of an entropic force acting on the mental state of the buyer/seller; and $\gamma$ can be interpreted as a kind of energy which is related to information (Orrell, 2020a).

As an example, suppose a propensity function $P(x)$ is normal with mean $\mu$ and standard deviation $\sigma$, where price $x$ is a logarithmic variable. Then the corresponding entropic force is

$$F(x) = \gamma \frac{P'(x)}{P(x)} = -k(x - \mu)$$

where $k = \gamma/\sigma^2$ is a force constant. The linear force, which is the same as that of a spring, therefore represents the mental desire for a buyer or seller to adjust the price to their own preferred level.

While the entropic force allows us to interpret the system in terms of dynamics, it doesn't tell us anything about the relevant mass that the force acts on; and as already seen we want to be able to account for effects such as interference and entanglement. We can address these issues by again moving to a quantum framework, and viewing the propensity function as being the product of an underlying wave function. The quantum version of a spring system is the quantum harmonic oscillator. This has a ground state whose corresponding probability distribution for position *x* is a normal distribution with mean $\mu$ and standard deviation $\sigma$. The associated mass is

$$m = \frac{\hbar}{2\omega\sigma^2}$$

so mass varies inversely with variance. Referring to Figure 4, the narrow propensity curve on the left has a higher associated mass than does the wider curve on the right. The scaling factor $\gamma$ is given by $\gamma = \hbar\omega/2$ which has units of energy.

In statistical physics, a frequency $\nu$ is related to the inverse of the Boltzmann time $\tau_B = \hbar/(k_B T)$ which is the theoretical order of time needed for an arbitrary nonstationary state to reach thermal equilibrium (Goldstein, Hara, and Tasaki, 2015). In the social version, the frequency can therefore be thought of as representing a linearised resistance to change. For something like a stock market, the frequency can be used to represent the speed of mean reversion of returns (Ahn et al, 2017).

In an economic transaction, the buyer propensity function will normally have a higher mean price than that of the seller, so the only parts of these curves which will be active are near the mid-price, as illustrated in Figure 5. It is easily checked that the joint propensity function, which is the product of the buyer and seller propensities, is a scaled normal curve and has an entropic force which is just the sum of the buyer and seller forces. The case where the supplier fixes the price and refuses to negotiate is handled by assuming an infinitely thin propensity curve located at the sale price.

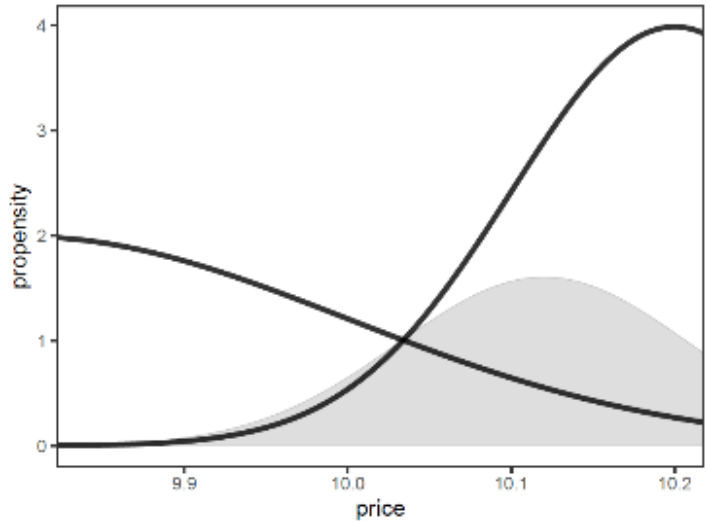

Figure 5. Propensity curves for a buyer (to the left) and a seller (to the right). The shaded area is the joint propensity.

Because the oscillator model is inherently probabilistic, it can be used to model features such as the statistics of financial markets, and the pricing and volume of financial options (Orrell, 2020a; 2021). Another difference between quantum and classical oscillators is that the former features discrete energy levels. The ground state, which again can be viewed as representing the potential for a transaction to occur, corresponds to the normal distribution, while the other states show more complicated distributions, and contribute the non-normal behavior also seen with markets (Ahn et al, 2017).

A main advantage of the quantum approach, when coupled with the entropic approach, is therefore that it gives us a consistent set of equations and units with which we can describe the dynamics of the system; and is particularly well adapted to the study of financial transactions, which involve the flow of information rather than just physical objects. For example, returning to the general case where the entropic force is given by

$$f(x) = \gamma \frac{P'(x)}{P(x)}$$

we can look at a particular mental state where the log price is $x_1$ and ask how much work – which again is linked to information – must be done against the entropic force to move to another state $x_2$. This is simply

$$\Delta E = \int_{x_1}^{x_2} F(x)\, dx = \gamma \log\left(\frac{P(x_2)}{P(x_1)}\right)$$

which depends only on the ratio of the initial and final propensities. The change in propensity required for preference reversal, as mentioned above, is typically a factor 3, which is close to Euler's number. It follows that what might be called the energy or information required to change a person's mind is

$$\Delta E = \frac{\hbar\omega}{2}\log(3) \cong \frac{\hbar\omega}{2}\log(e) = \frac{\hbar\omega}{2}$$

which is the base energy of a quantum oscillator. In a quantum computer, it is also the order of energy needed to flip a qubit from one state to another.

## 5. Conclusion

In early neoclassical economics, utility was viewed as a kind of energy. In his 1892 book *Mathematical Investigations in the Theory of Value and Prices*, Irving Fisher for example expressed economic transactions in physical terms, where utility had units of energy. The quantum framework returns to this idea of energy, but associates it with a change in propensity rather than a utility. While previous works have demonstrated the usefulness of quantum methods in modelling cognitive phenomena, the propensity approach allows us to view economic transactions in general as a quantized probabilistic system, of the sort simulated in quantum computing.

Of course, many social scientists will argue that it is inappropriate to quantify things like social power or mental energy, since they cannot be reduced to exact equations. However one consequence of the rational utility-optimizing picture is that complex social topics such as power relationships were downplayed or ignored (Häring and Douglas, 2013; Orrell, 2017).[3] Since economics is a quantitative discipline, we need a suitable framework with which to describe the interplay between objective and subjective forces that make up power.

---

[3] The topic of power receives more attention from heterodox economists. For example the "capital as power" approach of Nitzan and Bichler (2009) sees money and capital as manifestations of power, but does not attempt to quantify power directly.

A main contribution of the quantum approach is to connect power with propensity. While human motivations cannot be reduced to equations, it seems reasonable to quantify the propensity for a person to make a particular decision (indeed, this is the entire basis for fields such as behavioural economics). The entropic force (along with its associated energy) is merely another way of expressing a propensity function. The approach can be used to simulate a variety of economic phenomena, from cognitive interference in the decision-making process, to the price of financial options.

To summarise, neoclassical economics was marked by a switch from a labour theory of value, to one based on utility. Quantum economics makes a similar switch, by expressing value in terms of a propensity function. Adopting a quantum probabilistic framework allows us to incorporate cognitive effects such as interference and entanglement; express basic economic quantities such as forces of supply and demand in consistent units; and consider both subjective and objective factors on an equal footing.

**Appendix**

Each measurement in a non-computational basis is equivalent to applying a unitary matrix and then measuring in the computational basis (Nielsen, 2002). It follows that the circuit in Figure 2 is equivalent to Figure A.1, where $A = \begin{bmatrix} a_{11} & a_{12} \\ a_{21} & a_{22} \end{bmatrix}$ and $B = \begin{bmatrix} b_{11} & b_{12} \\ b_{21} & b_{22} \end{bmatrix}$ are unitary matrices determined based on the given non-computational basis. This Appendix shows how this one-qubit circuit is equivalent to the two-qubit quantum circuit depicted in Figure 3 for the general case where *A* and *B* are unitary matrices.

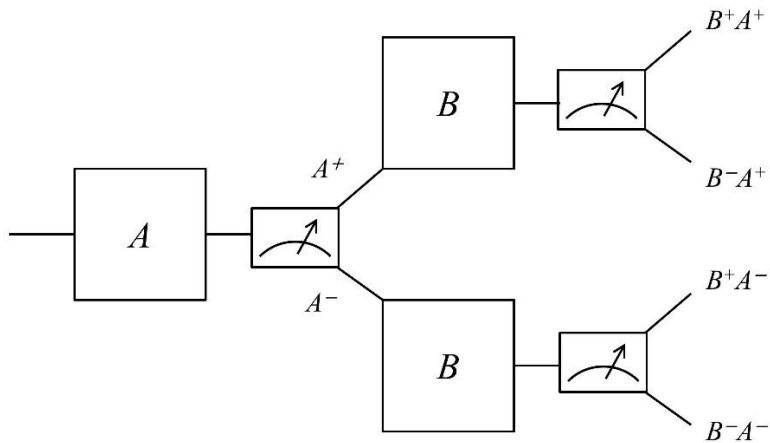

Figure A.1. A one-qubit circuit with two stages of measurement.

For the one-qubit circuit in Figure A.1, the qubit after gate $A$ will be in the state $A|0\rangle = \begin{bmatrix} a_{11} \\ a_{21} \end{bmatrix}$. The first measurement collapses the state to $|0\rangle$ with probability equal to $|a_{11}|^2$ (EVENT $A^+$) and to $|1\rangle$ with probability $|a_{21}|^2$ (EVENT $A^-$). For the order effect, $A^+$ would correspond to answering "Yes" to the first question, and $A^-$ would correspond to answering "No".

If EVENT $A^+$ happens, then the output state after applying $B$ is $B|0\rangle = \begin{bmatrix} b_{11} \\ b_{21} \end{bmatrix}$. The second measurement collapses the state to $|0\rangle$ with probability equal to $|b_{11}|^2$ (EVENT $B^+$) and collapses to $|1\rangle$ with probability equal to $|b_{21}|^2$ (EVENT $B^-$).

With the same argument for the other path we have:

| EVENTS | $A^+B^+$ | $A^+B^-$ | $A^-B^+$ | $A^-B^-$ |
|---|---|---|---|---|
| Result State | $|00\rangle$ | $|01\rangle$ | $|10\rangle$ | $|11\rangle$ |
| Probability | $|a_{11}b_{11}|^2$ | $|a_{11}b_{21}|^2$ | $|a_{21}b_{12}|^2$ | $|a_{21}b_{22}|^2$ |

On the other hand the output of the two-qubit circuit, when the control qubit is at the top of the circuit, is $a_{11}b_{11}|00\rangle + a_{11}b_{21}|01\rangle + a_{21}b_{21}|10\rangle + a_{21}b_{11}|11\rangle$ with associated probabilities:

| Result State | $|00\rangle$ | $|01\rangle$ | $|10\rangle$ | $|11\rangle$ |
|---|---|---|---|---|
| Probability | $|a_{11}b_{11}|^2$ | $|a_{11}b_{21}|^2$ | $|a_{21}b_{21}|^2$ | $|a_{21}b_{11}|^2$ |

Since *B* is unitary, its columns (or its rows) form an orthonormal basis (Steeb, 2006). In the case of a 2-by-2 unitary matrix, $|b_{11}|^2 + |b_{12}|^2 = |b_{11}|^2 + |b_{21}|^2 = 1$ which implies $|b_{12}|^2 = |b_{21}|^2$, and $|b_{11}|^2 + |b_{21}|^2 = |b_{21}|^2 + |b_{22}|^2 = 1$ which implies $|b_{11}|^2 = |b_{22}|^2$. It therefore follows that the final probabilities for the two circuits are the same. (Note that the intermediate probabilities are not the same.)